\pdfoutput=1
\documentclass[11pt]{article}
\usepackage[T1]{fontenc}
\usepackage[utf8]{inputenc}
\usepackage[a4paper]{geometry}
\usepackage[sc]{mathpazo}
\usepackage{amsmath,amssymb,mathtools,graphicx}
\usepackage{bm,soul}
\usepackage[cal=boondox]{mathalfa}
\usepackage{hyperref}
\hypersetup{hyperindex=false,pdftex}
\usepackage{dsfont,bbm}

\newcommand*\laplace{\mathop{\Delta}}

\begin{document}

\title{R. F\"urth's 1933 paper ``On certain relations between classical Statistics and Quantum Mechanics'' [``\"Uber einige Beziehungen zwischen klassischer Statistik und Quantenmechanik'', \textit{Zeitschrift für Physik,} \textbf{81} 143-162]}

\author{\textit{introduced by} Luca Peliti\\
Santa Marinella Research Institute\\
00058 Santa Marinella, Italy\\
\href{mailto:luca@peliti.org}{luca@peliti.org}\\[0.2cm]
\textit{translated and with a commentary by}\\
Paolo Muratore-Ginanneschi\\
Department of Mathematics and Statistics, University of Helsinki\\ 
P.O. Box 68, 00014 Helsinki, Finland\\
\href{mailto:paolo.muratore-ginanneschi@helsinki.fi}{paolo.muratore-ginanneschi@helsinki.fi}}

\date{\today}


\maketitle

\begin{center}
\textbf{Abstract}
\end{center}

\begin{quote}
We present a translation of the 1933 paper by R. Fürth in which a profound analogy between quantum fluctuations and Brownian motion is pointed out. This paper opened in some sense the way to the stochastic methods of quantization developed almost 30 years later by Edward Nelson and others.
\end{quote}

\section*{Introduction}
Thermodynamic uncertainty relations are a remarkable set of inequalities in Stochastic Thermodynamics that bound the coefficient of variation of empirical currents to their averages and to the entropy production rate (see, e.g., the research papers \cite{barato2015thermodynamic,GiRoHo17} and/or the monograph \cite{PePi20} for overview). In a nutshell, they intimate that achieving very accurate currents, with a very small coefficient of variation, requires in general a minimal cost in terms of entropy production. Their name are evocative of the uncertainty relations of quantum physics, which set a bound on the accuracy by which the position of a particle and its velocity can be evaluated. 

It is interesting to remark that a similar analogy between the inequality expressing Heisenberg uncertainty relations and a similar one which applies to diffusion processes like Brownian motion was pointed out in a remarkable paper by Reinhold Fürth in 1933. The paper, which points to a profound analogy between the uncertainty arising from quantum fluctuations and that due to random forces acting on a diffusing particle, opened the way in some sense to more recent developments like Nelson's stochastic mechanics approach~\cite{Nelson85} to quantum mechanics, and the stochastic quantization approach championed by Parisi and Wu~\cite{parisi1981perturbation,Nam92}.

We hope to be helpful to the community by providing a translation of this comparatively little known paper. The translation is preceded by a brief biographical skecth of its author and is followed by some remarks on the translation and on some early developments of the approach initiated by the paper. Notice that the author's references appear as footnotes as in the original paper. The references due to the curator appear in brackets and are listed at the end.

\section*{Reinhold Fürth}
Reinhold Fürth was born in Prague (then Austria-Hungary) in 1893. He obtained a doctorate from the German Charles-Ferdinand University in Prague in 1916, where he became a professor of experimental physics in 1931. After the German takeover in 1938-39 he moved to Scotland and became a research fellow at the University of Edinburgh. After having been elected fellow of the Royal Society in 1943, he moved in 1947 to Birkbeck College, London, to become there professor of theoretical physics. He died in 1979.

His name is well known as the editor of a collection of papers by Albert Einstein on the theory of Brownian movement~\cite{FurEin22}, whose English translation~\cite{FurEin56} is widely read. His lecture on the physics of social equilibrium~\cite{Fur51}, read before the British Association for the Advancement of Science in Edinburgh in 1951, can be considered as one of the earliest examples of the approach known as sociophysics.


\section*{Text}
\begin{center}
From the Physics Institute of the German University in Prague\\
\textbf{\large On certain relations between classical Statistics and Quantum Mechanics.}\\
From Reinhold Fürth in Prague.\\
With 4 figures. (Received on January 19, 1933.)
\end{center}

\begin{center}
\textbf{Abstract}
\end{center}
\begin{quote}
It is highlighted the formal analogy between the differential equations for the probability distribution of the position of a mechanical system according to classical Statistics and to Quantum Mechanics which can also  be interpreted  as equations for the motion of a cluster of identical particles, a diffusion. The physical origin of such diffusion will be ascribed in the classical case to the collision with molecules of the surrounding  matter,
in the case of Quantum Mechanics to the uncertainty relations. In the last case, diffusion in the absence of forces is discussed and a simple derivation of the uncertainty relations is given on this basis. The line of reasoning can be carried over to classical diffusion and it is
possible to derive an inequality for the variance of the position and the velocity which is in strict analogy with \so{Heisenberg}'s uncertainty relations. The  relation found can be also applied to a single particle and more generally to an arbitrary mechanical system, since it states that the simultaneous measurement of the position and of the corresponding velocity is possible only up to a maximal accuracy in consequence of the \so{Brown}ian motion. It is discussed the relation of this finding with the problem of  [determining] with which accuracy it is possible to measure a physical quantity with a mechanical measurement device and as a result it turns out that there exists also here in analogy [with Quantum Mechanics] an accuracy limit which cannot be overcome.
Finally, it is shed light from the point of view of wave Mechanics on the question why the classical diffusion equation holds for a real density function with a real diffusion coefficient in contrast to the Schr\"odinger equation [which holds] for a complex function with an imaginary coefficient, and [this fact] is related to the problem of the observability of physical quantities and of the reversibility versus irreversibility of natural processes.  
\end{quote}


\section*{}
\label{sec:intro}

In what follows there shall be a discussion of certain relations between classical statistics  -- the classical diffusion theory and the theory of \so{Brown}ian motion -- on the one hand and,  Quantum Mechanics  on the other, [discussion] which arises from formal reasons and [which], although it might be already known to some, to the best of my knowledge has yet not been addressed in this context. In particular it is possible to show that \so{Heisenberg}'s uncertainty relations carry over to processes which are governed by classical Statistics and that  it is thus possible to bring about new perspectives on the often addressed question of the limit of  measurability with an measurement device. It is furthermore attempted to make precise the physical meaning of the aforementioned similarities and differences.

\section{}\label{sec:1} The classical theory of diffusion is governed~\footnote{See, e.g., Frank-Mises, \textsl{Differential-u. Integralgleichungen d. math. Physik} 2, 248} 
by the generalised diffusion equation
\begin{align}
\frac{\partial u}{\partial t}= D \,\laplace u-\operatorname{div}(u \mathfrak{v})
\label{diffeq}
\end{align}
where $ u(x,y,z,t) $ denotes the density as function of the position and time, $ D $ (assumed constant) the diffusion coefficient and $ \mathfrak{v} $ the velocity vector of the convection current occasioned by external forces. The solution of this equation under given boundary conditions determines the distribution of the density at any future instant of time if the distribution is known in the present.

If one interpret the diffusion experiment as a collective experiment with a spatial ensemble of many identical particles then $ u \mathrm{d}V $ is the relative frequency with which any element of the ensemble is found  in the volume element $ \mathrm{d}V $ at time $ t $ during the collective experiment  if $ u $ satisfies the normalization condition
\begin{align}
\iiint\,u\,\mathrm{d}V=1
\label{norm}
\end{align}       
for all $ t $. The replacement of the spatial ensemble with a virtual ensemble turns the diffusion equation (\ref{diffeq}) into an equation for the ``probability density'' $ u $ of the position of an individual particle which can be computed as a function of time when it is known at time zero: \so{Smoluchowski}'s differential equation for \so{Brown}ian motion of an individual particle under the action of external forces~\footnote{M. v. Smoluchowski, \textit{Ann. d. Phys.} 43, 1105, 1915}.

It is possible to show that \so{Smoluchowski}'s equation is a special case of another differential equation which can be derived under very general conditions for the \so{Brown}ian motion of an arbitrary mechanical system and [which] is usually referred to as the \so{Fokker-Planck} equation~\footnote{See, among others, F. Zernike, \textsl{Handb. d. Phys. Bd. III,} S. 457}. 
Following \so{Schr\"odinger}~\footnote{E. Schr{\"{o}}dinger, \textit{Ann. de l'Inst. H. Poincar\'e} 1931, S. 296ff. \textit{S. Ber. Berl. Akad.} 1931, S. 148; see also J. Metadier, \textit{C. R.} \textbf{193,} 1173, 1931.}, [the Fokker-Planck equation]  can be written as
\begin{align}
\frac{\partial u}{\partial t}= F u
\label{FPeq}
\end{align}
where $ F $ denotes a certain differential operator which, in agreement with (\ref{diffeq}), reduces to $ F=D \Delta- \operatorname{div}\mathfrak{v} $  in the case when the system is a particle under the action of a force.

The differential equation (\ref{FPeq}) is, as \so{Schr\"odinger}~\footnote{E. Schr{\"{o}}dinger, \textit{Ann. de l'Inst. H. Poincar\'e} 1931, S. 296ff. \textit{S. Ber. Berl. Akad.} 1931, S. 148; see also J. Metadier, \textit{C. R.} \textbf{193,} 1173, 1931. [sic]} 
also already pointed out, formally identical to the time dependent \so{Schr\"odinger} differential equation of wave mechanics for the wave function $ \psi $ which is usually written
in the form
\begin{align}
\label{Seq}
\frac{\partial \psi}{\partial t}= H \psi
\end{align}
where $ H $ denotes the \so{Hamilton} operator for the mechanical problem of interest.
According to the statistical formulation of wave mechanics also this equation is a 
``probability equation'' inasmuch it allows one to compute this quantity at any arbitrary later instant of time from the knowledge of $ \psi(q) $ at time zero and the ``probability amplitude'' $ \psi $ is linked to the probability density for the sojourn of the system in a certain volume element of the $ q $-space by the relation
\begin{align}
\label{Born}
w=\psi\psi^{*}
\end{align}
($ \psi^{*} $ is the complex conjugate of $ \psi $) as far as $ \psi $ satisfies the
normalization condition
\begin{align}
\iiint\,\psi\psi^{*}\,\mathrm{d}V=1
\label{psinorm}
\end{align}
By reversing the line of reasoning,  one can also construe the quantity $ w $ defined via (\ref{Born}) as the phase point density of a large number of identical non interacting systems in $ q $-space. Equation
(\ref{Seq}) then determines the evolution of such distribution density and permits to compute the density at any further time if the density function is assigned at time zero. 

In the case of special importance  of an individual point system of mass $ m $ subject to the action of a force which can be derived from a potential $ U $, equation (\ref{Seq}) reads
\begin{align}
\label{}
-\frac{h}{2\,\pi\, i }\frac{\partial \psi}{\partial t}=-\frac{h^{2}}{8\,\pi^{2}\,m}\laplace\psi+U\,\psi
\end{align}
The discussion of this equation teaches, as \so{Ehrenfest} first showed~\footnote{P. Ehrenfest, \textit{ZS. f. Phys.}\textbf{45}, 455, 1927}
 that the centre of mass of a cluster of particles obeying the conditions expounded afore 
moves in the usual three dimensional space  according to the prescription of classical mechanics when  the assigned forces act on the particle but also that the cluster of particles spreads around the centre of mass via  a sort of diffusion. We therefore encounter here  a convection current with overlaid a diffusion in analogy with the motion of a cluster of particles according to the classical theory of diffusion.

As we are interested only in the last phenomenon, we wish in what follows set to zero the external force. The equations (\ref{Seq}) and (\ref{diffeq}) become then formally identical, namely
\begin{align}
\label{freediff}
\frac{\partial u}{\partial t} = D\, \laplace\,u
\end{align} 
and 
\begin{align}
\label{freeS}
\frac{\partial \psi}{\partial t}=\epsilon \laplace\,\psi
\end{align} 
where the shortcut 
\begin{align}
\label{sc}
\epsilon=\frac{ i \,h}{4\,\pi\,m}
\end{align}
is used. Subject to the same boundary and initial conditions the solutions of (\ref{freediff}) and (\ref{freeS}) read hence completely the same. Sure enough a substantial difference arises from the fact that
in the case of Quantum Mechanics not the function (in general complex) $ \psi $ but rather according to (\ref{Born}) its norm plays the role of density function and that according to (\ref{sc}) the diffusion coefficient is here purely imaginary.  We return to the physical meaning of this fact later below.

\section{}\label{sec:2}
The deeper reason for the analogy emerging in the comparative presentation of \S~1 between the motion of a cluster of particles according to the classical theory of diffusion and Quantum Mechanics resides in the fact that in both cases the velocities of  individual particles in the cluster differ and obey  a statistical law.

In the first case, this (phenomenon) stems from the fact that particles incur in irregular collisions with molecules of the surrounding element whereby the particles' momentum is continuously varied in intensity and direction in such a way that there is no relation between the change of momenta of  distinct particles. This [fact] becomes manifest when considering  an individual particle in its irregular \so{Brown}ian motion,
and, when considering a particle cluster, in the fact that for an assigned initial state of the cluster and initially vanishing ``macroscopically'' measured velocity, the particles actually possess velocity irregularly distributed across the cluster and that in the course of time the initial distribution varies in the characteristic way of a diffusion. 

In the case of Quantum Mechanics, the very assumption of an initial density distribution  implies that the condition of vanishing initial velocity of all the particles cannot be strictly satisfied. According to \so{Heisenberg}'s
 fundamental uncertainty relations governing Quantum Mechanics a complete assignment of the initial velocity of the particles would be possible only in the presence of a complete uncertainty about the initial positions.

As a certain information about the initial position of the particles is conveyed by the assignement of the initial distribution, one must admit a certain blurring of the initial velocities, i.e., a certain statistical distribution of the initial velocities of the cluster particles. But a necessary consequence of this is that  a variation of the initial density distribution as well as a diffusion of the cluster must have occurred after a certain time.

That the uncertainty  on the value of position of the particles of the diffusing cluster really satisfies \so{Heisenberg}'s uncertainty relations with the uncertainty about the value  of the velocity (momentum), has been shown by \so{Heisenberg}~\footnote{W. Heisenberg, \textit{ZS. f. Phys.} \textbf{43,} 172, 1927.}
and \so{Kennard}~\footnote{E. H. Kennard, {loc. cit.} \textbf{44,} 326, 1927.} 
among others. A brief derivation may be given here for the one-dimensional case,
which, without resorting to the theory of transformations, makes use~\footnote{I need to thank here Mr. K. \so{L{\"o}wner}, Prague, for some hints.} only of equation (\ref{freeS}) and its complex conjugate taking in one dimension the form
\begin{align}
\left. 
\begin{array}{l}
\dfrac{\partial \psi}{\partial t} = \epsilon \dfrac{\partial^{2} \psi}{\partial x^{2}}
\\[0.3cm]
\dfrac{\partial \psi^{*}}{\partial t} = -\epsilon \dfrac{\partial^{2} \psi^{*}}{\partial x^{2}}
\end{array}
\right  \}
\label{system}
\end{align}
Let $ x_{0} $ the initial position of one particle of the cluster, $ v $ its initial velocity and $ x $ its position after a time $ t $, then 
\begin{align}
\label{linear}
x=x_{0}+v\,t
\end{align}
holds.
If the centre of mass at time zero is located in the origin of the coordinates and its velocity is zero, i.e. $ x_{0}=0 $ and $ \bar{v}=0 $ then according to (\ref{linear}) it is also clear that 
$ \bar{x}=0 $ for all $ t $. Evaluating the quadratic expectation value of  (\ref{linear}) one gets into
\begin{align}
\label{quadratic}
\overline{x^{2}}=\overline{x_{0}^{2}}+2\,\overline{x_{0}\,v}\,t+\overline{v^{2}}\,t^{2}
\end{align}
 By definition 
 \begin{align}
 \overline{x^{2}}=\int_{-\infty}^{+\infty}x^{2}\psi\psi^{*}\mathrm{d}x
 \label{defx2}
 \end{align}
 holds true.
 Using equation (\ref{system}) and under the assumption that $ \psi $ vanishes sufficiently fast at infinity, one gets into after a simple calculation 
 \begin{align}
 \label{der1}
 \frac{\mathrm{d}}{\mathrm{d} t}\overline{x^{2}}=2\,\epsilon\,\int_{-\infty}^{\infty} x \left(\psi\frac{\partial \psi^{*}}{\partial x }-\psi^{*}\frac{\partial \psi}{\partial x }\right)\mathrm{d}x
 \end{align}
 \begin{align}
 \label{der2}
 \frac{\mathrm{d}^{2}}{\mathrm{d} t^{2}}\overline{x^{2}}=-8\,\epsilon^{2}\,\int_{-\infty}^{\infty} \frac{\partial \psi^{*}}{\partial x }\frac{\partial \psi}{\partial x }\mathrm{d}x
 \end{align}
 \begin{align}
 \label{der3}
 \frac{\mathrm{d}^{3}}{\mathrm{d} t^{3}}\overline{x^{2}}=0
 \end{align}
 From (\ref{der3}) it follows that $  \overline{x^{2}}$ must be a quadratic function of time in agreement with (\ref{quadratic})  ;   it  also  follows from (\ref{der2}) that $ \overline{v^{2}} $ as coefficient of  $ t^{2} $  (\ref{quadratic}) satisfies
 \begin{align}
 \label{v2}
 \frac{\mathrm{d}^{2}}{\mathrm{d} t^{2}}\overline{v^{2}}=\frac{1}{2}\frac{\mathrm{d}^{2}}{\mathrm{d} t^{2}}\overline{x^{2}}=-\,4\,\epsilon^{2}\,\int_{-\infty}^{\infty} 
 \left |\frac{\partial \psi}{\partial x }\right |^{2}\mathrm{d}x
 \end{align}
 According to \so{Heisenberg}~\footnote{W. \so{Heisenberg}, \textsl{Die physikalischen Prinzipien der Quantentheorie.}	Leipzig 1930. Page 13 and following.}, 
 it now follows from the self-evident inequality
 \begin{align}
 \label{ineq1}
 \left |\frac{x}{2\,\bar{x}}\psi+\frac{\partial \psi}{\partial x}\right |^{2}\,\geq\,0
 \end{align}
 with use of (\ref{norm}) and (\ref{defx2})
 \begin{align}
 \nonumber
 \int_{-\infty}^{\infty} 
 \left |\frac{\partial \psi}{\partial x }\right |^{2}\mathrm{d}x\,\geq\,\frac{1}{4\,\overline{x^{2}}}
 \end{align}
 whence from (\ref{der3})
 \begin{align}
 \label{Heisenberg1}
 \overline{x^{2}}\overline{v^{2}}\,\geq\,-\epsilon^{2}
 \end{align}
If one introduces the uncertainties on the position and momentum
of the particle cluster under consideration with the avail of the relations
\begin{align}
\label{uncertain}
\left .
\begin{array}{l}
\Delta x = \sqrt{\overline{x^{2}}}
\\[0.3cm]
\Delta p = m \sqrt{\overline{v^{2}}}
\end{array}
\right \}
\end{align} 
using (\ref{sc}) from (\ref{Heisenberg1}) follows for their product \so{Heisenberg}'s relation
\begin{align}
\Delta x \, \Delta p \,\geq\, \frac{h}{4\,\pi}
\label{Heisenberg}
\end{align}
The equality holds here if and only if the inequality (\ref{ineq1}) turns in an equation.
The integration of the latter yields for $ \psi $
\begin{align}
\psi=\mathrm{Const.} \,e^{-x^{2}/4\,(\Delta x)^{2}}
\label{Gpsi}
\end{align}
and also for the density of the particle cluster (\ref{Born}) in consideration of (\ref{norm}) the Gaussian distribution
\begin{align}
\label{Gauss}
w=\frac{1}{\sqrt{2\,\pi}\Delta x}\,e^{-x^{2}/2\,(\Delta x)^{2}}
\end{align}
If $ \psi $ takes at time $ t=0 $ the special form (\ref{Gpsi}) then it follows from 
(\ref{der1}) $ \frac{\mathrm{d}}{\mathrm{d} t} \overline{x^{2}}=0$ and as a consequence the coefficient of $ t $ disappears from (\ref{quadratic}). If there is a corresponding initial distribution of the position in the particle cluster under consideration, so that $ (\overline{x_{0} v})=0 $ [holds] at the same time, the variance of the positions and of the initial velocities of the single particles are also statistically independent one from another. Conversely, by no means the existence at time zero of the density distribution (\ref{Gauss}) implies by itself the statistical independence of position and velocities and hence the vanishing of the linear term in (\ref{quadratic}).

\section{}\label{sec:3}
In accordance with what was said at the beginning of section~\ref{sec:2}, it is natural to apply the above reasoning, which is based on the \so{Heisenberg} uncertainty relation in the quantum mechanical case, to the case of classic diffusion. Also in this case we restrict ourselves to the one dimensional case with vanishing convection current and so we start from equation (\ref{freediff}) which in one dimension reads
\begin{align}
\label{3:freediff}
\frac{\partial u}{\partial t}=D \frac{\partial^{2} u}{\partial x^{2}}
\end{align}
where in agreement with (\ref{norm}) $ u $ satisfies the condition
\begin{align}
\int_{-\infty}^{\infty} u\,\mathrm{d}x=1
\label{3:norm}
\end{align}
We define the uncertainty of the particle cluster by means of the quantity $ \overline{x^{2}} $ as
\begin{align}
 \overline{x^{2}}=\int_{-\infty}^{\infty}x^{2}\, u\,\mathrm{d}x
\label{3:var}
\end{align}
At $ t=0 $ the centre of mass of the cluster lies again in the origin of the coordinates so that $ \overline{x_{0}}=0 $. To start with, we look for the derivation of  the analog of equation (\ref{quadratic})
which expresses how the uncertainty $ \overline{x^{2}} $ initially present in the diffusing particle cluster grows in the course of time.
\begin{align}
\frac{\mathrm{d}}{\mathrm{d} t}\overline{x}=\frac{\mathrm{d}}{\mathrm{d} t}\int_{-\infty}^{\infty}x\, u\,\mathrm{d}x
=
\int_{-\infty}^{\infty}x\, \frac{\partial u}{\partial t}\,\mathrm{d}x\\
=
D\int_{-\infty}^{\infty}x\, \frac{\partial^{2} u}{\partial x^{2}}\,\mathrm{d}x=0
\nonumber
\end{align}
The center of mass of the cluster constantly remains at rest, as the absence of a convection  immediately evinces, so that for all times
$ \overline{x}=0 $.

From (\ref{3:var}) in an analogous way it follows that
\begin{align}
\label{3:dvar}
\frac{\mathrm{d}}{\mathrm{d} t}\overline{x^{2}}=
\int_{-\infty}^{\infty}x^{2}\, \frac{\partial u}{\partial t}\,\mathrm{d}x
=
D\int_{-\infty}^{\infty}x^{2}\, \frac{\partial^{2} u}{\partial x^{2}}\,\mathrm{d}x= 2\,D
\end{align}
and therefore that $ \overline{x^{2}} $ is a linear function of time of the form
\begin{align}
\overline{x^{2}}=\overline{x_{0}^{2}}+2\,D\,t
\label{3:quadratic}
\end{align}
The comparison of (\ref{3:quadratic}) with (\ref{quadratic}) shows that  by all means  in both cases the uncertainty over the position indefinitely grows over a sufficient long time and thus that a diffusion of the cluster occurs.  
Whereas here, however, the growth of $ \overline{x^{2}} $ occurs independently of $ \overline{x_{0}^{2}} $ and \textit{linearly} in time, there the growth in time is \textit{quadratic} and in consequence of (\ref{Heisenberg1}) is itself dependent upon $ \overline{x_{0}^{2}} $ (it takes place in a particularly sudden way if $ \overline{x_{0}^{2}} =0$ inasmuch $ \overline{v^{2}} $ becomes infinitely large); finally, if the linear term in $ t $ is non-vanishing the so that dispersion of the positions and of the velocities are not statistically independent at time zero, it may be that the cluster undergoes first a contraction to a minimum and only afterwards a spreading.

The formal causes for the aforementioned  differences have been already discussed at the end of \S~1. The differences are physically explained by the fact that in the case of classical diffusion there is no ``initial velocity'' of the particles and therefore no equation of the form (\ref{linear}) exist, and that furthermore the instantaneous speed of the particles is due to the collisions with the molecules, as already mentioned~\footnote{If one, following \so{Schr{\"o}dinger} (Ber. Ber. 1930, S. 296, Nr. 19), sets out to choose the value of $ \overline{x_{0}^{2}} $ such that
$ \Delta x $ is as small as possible and the product $ \Delta x \Delta p =\frac{h}{4\,\pi} $ holds true so that the initial distribution (\ref{Gpsi})
is satisfied, then in (\ref{quadratic}) the second term of the right hand side vanishes whereas the first and the third become equal upon setting $ \overline{v^{2}} = - \frac{\epsilon^{2}}{\overline{x^{2}}}$. This implies for $ \overline{x^{2}} $ 
\begin{align}
\overline{x^{2}}=\frac{h}{2\,\pi\,m}t
\label{3:note}
\end{align}
In this case, the formal analogy with (\ref{3:quadratic}) is strikingly evident when one thereby replaces $ D $ with the absolute value of the  ``imaginary diffusion coefficient'' $ \epsilon $ according to (\ref{sc}). }.
On the basis of the statistical independence of the dispersion process and the initial distribution in the classic case, one can immediately write equation (\ref{3:quadratic}), since it expresses that this is due to two causes: the ``square error'' of $ x $ resulting from  initial spread and diffusion is the sum of these two just mentioned ingredients, the second being the well-known \so{Einstein} law for the mean square of the \so{Brown}ian motion.

In order to find the analog of the uncertainty relations (\ref{Heisenberg1}) to (\ref{Heisenberg}), we need in first place to find a suitable definition of velocity for the classical diffusion. From the above it is clear that by no means this role can be played by the microscopic velocity produced by molecular collisions. Likewise, as we have already seen, the macroscopic velocity of the cluster regarded as a single entity, or strictly speaking the velocity of its centre of mass, [is not a good candidate since it] vanishes. A suitable quantity comes about from the consideration of
the ``diffusion current'', i.e., the quantity of diffusing matter crossing in the unit of time a fixed section in the diffusion domain.
As it is well known~\footnote{Compare ref.~1.}, the vector $ \mathfrak{Q} $ of the diffusion current is a local function in the diffusion domain, connected with the scalar $ u $ by the relation
\begin{align}
\label{3:j}
\mathfrak{Q} = - D \operatorname{grad}u
\end{align} 
Based on the fact that $ u $ is nothing else than density of matter of the diffusing element, 
we find the corresponding velocity vector $ \mathfrak{v} $ according to
\begin{align}
\mathfrak{v}=\frac{1}{u}\mathfrak{Q}=- D \frac{1}{u}\operatorname{grad}u
\label{3:cv}
\end{align}
which in the one-dimensional case becomes
\begin{align}
\label{3:cv1d}
v=- D \frac{1}{u}\frac{\partial u}{\partial x}
\end{align}
If we now compute the particle cluster mean value of $ v $ at a certain time instant, we obtain by 
definition using (\ref{3:freediff})   
\begin{align}
\label{3:meancv}
\overline{v}=\int_{-\infty}^{\infty}\,v\, u\, \mathrm{d}x =- D\int_{-\infty}^{\infty}\frac{\partial u}{\partial x} \mathrm{d}x =0
\end{align}
as it must be, since $ \overline{v} $ is nothing else than the macroscopic velocity of the centre of mass.

For the mean value of $ \overline{v^{2}} $, one finds
\begin{align}
\label{3:cvvar}
\overline{v^{2}}=\int_{-\infty}^{\infty}v^{2} \,u \,\mathrm{d}x= D^{2}\int_{-\infty}^{\infty} \frac{1}{u}\left(\frac{\partial u}{\partial x}\right)^{2}\mathrm{d}x
\end{align}
By a straightforward application of the reasoning of \S~2 one can establish the derivation of an inequality for the product $ \overline{v^{2}}\overline{x^{2}} $, by proceeding once again from the self-evident inequality
\begin{align}
\label{3:obvious}
\left(\frac{1}{u}\frac{\partial u}{\partial x}+\frac{x}{\overline{x^{2}}}\right)^{2}\,\geq\,0
\end{align}
whence by expanding the product it follows
\begin{align}
\frac{1}{u}\left(\frac{\partial u}{\partial x}\right)^{2}\,\geq\,-2\frac{x}{\overline{x^2}}\frac{\partial u}{\partial x}-\frac{x^{2}\,u}{(\overline{x^2})^{2}}
\nonumber
\end{align}
Upon integrating, a simple calculation making use of (\ref{3:norm}) and (\ref{3:var}) yields
\begin{align}
\nonumber
\int_{-\infty}^{\infty}\frac{1}{u}\left(\frac{\partial u}{\partial x}\right)^{2}\,\geq\,\frac{1}{\overline{x^2}}
\end{align}
whence finally according to (\ref{3:cvvar})
\begin{align}
\overline{x^{2}}\,\,\overline{v^{2}}\,\geq\,D^{2}
\label{fuerth}
\end{align}
As one can see, the inequality (\ref{fuerth}) has the same form of the inequality (\ref{Heisenberg1}), which turns into (\ref{fuerth}) if one again replace the absolute value of $ \epsilon $ with $ D $.
Introducing the notation $ \Delta x $ and $ \Delta p $ in analogy with  (\ref{uncertain}), we write our uncertainty relation in the simpler form
\begin{align}
\label{3:main}
\Delta x \, \Delta p \,\geq\, D
\end{align} 
stating that in a classically diffusing particle cluster the position and the velocity of the particles at any instant of time cannot be \textit{simultaneously} determined with arbitrary accuracy and that furthermore the product of the uncertainties must be always larger than the diffusion coefficient $ D $.

The lower bound is attained, i.e., the inequality turns into an equality if and only if (\ref{3:obvious}) [also] holds as an equality. The solution of the differential equation obtained in this way immediately yields 
\begin{align}
u=\frac{1}{\sqrt{2\,\pi}\Delta x}\,e^{-x^{2}/2(\Delta x)^{2}}
\label{3:Gauss}
\end{align}
having taken (\ref{3:norm}) into account, and [is] therefore again the \so{Gauss}ian distribution, as in the quantum mechanical case, in formal agreement with (\ref{Gauss}).

Whereas in the present case, from the occurrence of the distribution (\ref{3:Gauss})  the  equality $ \Delta x \, \Delta p \,=\, 0 $ necessarily follows, the occurrence of the distribution (\ref{Gauss}) is there only a necessary but not sufficient condition for the product   $ \Delta x \, \Delta p $ to attain its minimum. Furthermore, whereas in a cluster of particles left to itself and satisfying at time zero the minimum uncertainty condition this condition continues to hold at any further time (because the distribution (\ref{3:Gauss}) is self-sustaining) in the quantum mechanical case the minimum condition is only instantaneously satisfied, e.g., at time zero and later no more (since the form of the distribution (\ref{Gpsi}) is not preserved by the motion of the particles). Finally, it should be emphasized that in the classical case one can always think a cluster of particles satisfying the minimum condition as one brought about by the diffusion of one which at a certain instant of time was completely concentrated in the origin of the coordinates. In order to see this,  one needs only to make use  in (\ref{3:Gauss}) of  (\ref{3:quadratic}) where one insert the abbreviation $ \overline{x_{0}^{2}}=2\,D\,t_{0} $;    one then obtains
\begin{align}
u=\frac{1}{2\sqrt{\pi \,D\,(t+t_{0})}}\,e^{-x^{2}/4\,D\,(t+t_{0})}
\label{3:diff}
\end{align}
which entails that indeed, for  $ t=-t_{0} $, $ u $  vanishes in the full space with the exception of $ x=0 $. In the quantum mechanical case this reduction, as we have already seen, is not possible.

\section{}\label{sec:4}
In the two preceding paragraphs we discussed the application of  uncertainty relations to a spatial aggregate of identical particles in the quantum and in the classical case. As it is well known, the fundamental significance of the uncertainty relation in Quantum Mechanics appears, however, when it is applied to an individual system. It teaches that the simultaneous measurement of the position and the momentum of a force-free particle can be performed with the maximum accuracy $ h/4\pi $ predicted by formula (\ref{Heisenberg}) since the measurement process  during the measurement of one of the two quantities disturbs the other to an amount that  the product of the uncertainties of both quantities cannot be lower than the aforementioned value. One can reformulate the statement for a general mechanical system by saying that the simultaneous measurement of a coordinate $q$ and of the impulse canonically conjugated to it is only possible with an uncertainty of the order of magnitude of $ h $. 

We can now also in a straightforward way apply the relation (\ref{3:main}) obtained in \S~3 to the problem of the simultaneous measurement of the position and speed of a particle, which is under the action of irregular impacts, and therefore performs a \so{Brown}ian motion.
Our relation teaches that the product of the uncertainty of a simultaneous measurement of position and velocity cannot be lower than the value $ D $, whereby  velocity must be understood as the macroscopic speed of the particle, i.e. the quantity $ \delta x /\delta t $ (assuming that $ \delta t $ is large compared to the time between two successive molecular collisions of the particle).
One sees that, as in the quantum mechanical case, there is an actual impossibility of a simultaneous, precise measurement of position and velocity, which, however, is not, as in Quantum Mechanics, determined by the process of measurement itself and governed by a universal constant, but it is rather caused by the influence of the environment on the observed system, and as a consequence it is clearly not of universal nature (for example, by lowering the temperature, which determines the value $ D $, [the effect] can be made arbitrarily small).

The following argument evinces that formula (\ref{3:main}) holds true also in the case of the measurement of an individual particle: we consider a force-free particle which at time zero is located at the origin of the coordinates and has vanishing macroscopic velocity. If we measure the position of the particle after a short time $ t $ then the expected value $ \overline{x^{2}} $ satisfies Einstein's formula
\begin{align}
\overline{x^{2}}=2\,D\,t
\label{4:Einstein}
\end{align} 
whence it follows
\begin{align}
\frac{\mathrm{d}}{\mathrm{d} t}\frac{\overline{x^{2}}}{2}=D
\nonumber
\end{align}
If we now exchange the order between time differentiation ad expectation value, we get furthermore
\begin{align}
\label{4:Stratonovich}
\overline{\left(\frac{\mathrm{d}}{\mathrm{d} t}\frac{x^{2}}{2}\right)}
=
\overline{\left(x\frac{\mathrm{d}}{\mathrm{d} t}x\right)}=D
\end{align}

$ x $ is evidently now the uncertainty over the position of the particle (we assumed  $ x=0 $ at time zero) caused by the \so{Brown}ian motion, and  
similarly $ \mathrm{d} x/\mathrm{d} t $ is the uncertainty over the velocity (which we assumed vanishing at time zero) brought about by the very same causes. The product $  \overline{\left(x\frac{\mathrm{d}}{\mathrm{d} t}x\right)}$ thus specifies the value to be expected by averaging over many measurements of the uncertainty product $ \Delta x \Delta v$ which  according to equation (\ref{4:Stratonovich}) is equal to $ D $.
The fact that we obtained exactly the minimum value here instead of equation (\ref{3:main}) is due to the fact we evaluated the mean value over repeated measurements of a particle, which we assumed to have always the same starting position and starting velocity at time zero. It is immediately obvious that without this assumption the uncertainty can in any case only increase, so that the product $ \Delta x \Delta v $ is actually larger than $ D $, as required by the relationship (\ref{3:main}).

Our relation states that an increase in the measurement accuracy of the position of a \so{Brown}ian particle reduces the accuracy of a simultaneous 
measurement of the velocity and vice versa. The physical meaning of this statement can be visualized with the help of the following Figs 1-4 of which Fig 1 plots as the function $ x(t) $ the position as a function of time of a particle falling under the effect of gravity in a liquid, observed with a certain magnification, and Fig. 2 represents the function $ v(t)=\dot{x}(t) $ obtained from it. Fig 3 shows the beginning of Fig. 1, plotted with a stronger magnification, and Fig. 4 again the velocity curve obtained therefrom.

\begin{figure}[htb]
\begin{center}
\includegraphics[width=0.4\textwidth]{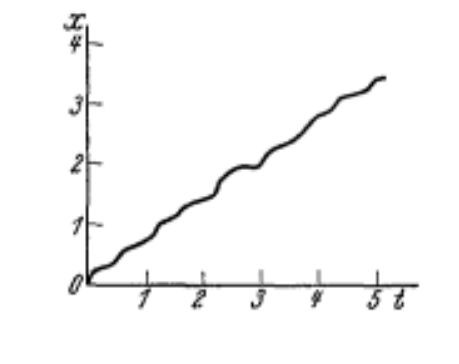}
\end{center}
\caption{Plot of the position of a \so{Brown}ian particle as a function of time (stylized).}
\end{figure}

\begin{figure}[htb]
\begin{center}
\includegraphics[width=0.4\textwidth]{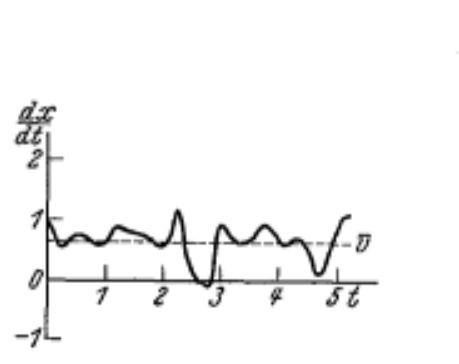}
\end{center}
\caption{Velocity $ v $ of the particle as a function of time computed from Fig.1 (dashed line mean value $ \overline{v} $). }
\end{figure}

\begin{figure}[htb]
\begin{center}
\includegraphics[width=0.4\textwidth]{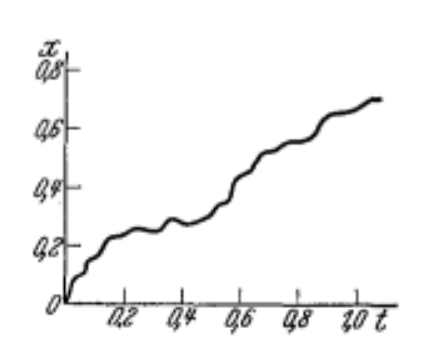}
\end{center}
\caption{5-times magnification of the beginning of the plot in Fig.~1 (stylized).}
\end{figure}

\begin{figure}[htb]
\begin{center}
\includegraphics[width=0.4\textwidth]{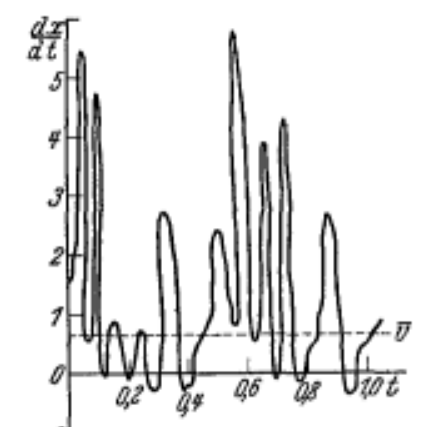}
\end{center}
\caption{Velocity $ v $  computed from Fig.~3 (dashed line mean value $ \overline{v} $).}
\end{figure}

One can immediately see how increasing the accuracy in the determination of the position by increasing the magnification necessarily increases the uncertainty in the simultaneous determination of the velocity.
Our relation thus expresses in an exact way the fact known to everyone familiar with \so{Brown}ian motion that the trajectory of a \so{Brown}ian particle exhibits more discontinuities with increasing magnification.

Exactly as in the case of Quantum Mechanics, we can extend the uncertainty relation (\ref{3:main}) also to any mechanical system in contact with a surrounding temperature bath. Then, to every degrees of freedom is associated the \so{Brown}ian motion of the corresponding coordinate which we denote again by $ x $. The \so{Fokker-Planck} equation (\ref{FPeq}) takes the place of the differential equations (\ref{3:freediff}) or (\ref{freediff}).
It is plausible that also in this general case an uncertainty relation on the form
\begin{align}
\Delta x\, \Delta v \approx D
\label{4:general}
\end{align}
holds true where $ v $ is the velocity associated to the coordinate $ x $, and $ D $ denotes the coefficient of the term $ \frac{\partial^{2} u}{\partial x^{2}} $ on the right hand side of (\ref{freediff}) and expresses the characteristic constant of this \so{Brown}ian motion. The relation
 states that the simultaneous measurement of the coordinate $ x $ and of its associated speed $ v $ is possible only with an uncertainty of order~$ D $.
 
 \section{}\label{sec:5}
 We can also extend the domain of validity of formula (\ref{4:general}) to any non-mechanical quantity since any physical quantity, even of non-mechanical nature, is measured using mechanical  measurement instruments, for example a current [is measured] using a galvanometer itself consisting of mechanical components. We assume that the ``deflection'' $ x $ of the  mechanical instrument in use be proportional to the  quantity $ J $ to be measured (for example the deflection of a galvanometer [is proportional] to the intensity of the current). When this is not the case from the start, one can always apply a compensation method in order to implement the desired condition within strict accuracy. Let $ \dot{J} $ be the speed of variation of $ J $. Then it holds true that
 \begin{equation}
 \begin{split}
 &J=a\,x\hspace{1.7cm}\dot{J}=a\,\dot{x}=a\,v
 \\
& \Delta J=a\,\Delta x\hspace{1.0cm}\Delta\dot{J}=a\,\Delta\dot{x}=a\,\Delta v
\end{split}
 \end{equation}
 whence with the help of (\ref{4:general}) 
 \begin{align}
 \Delta J\,\Delta\dot{J}\approx a^{2}\,D
 \label{5:general}
 \end{align}
 The relation  (\ref{5:general}) teaches that although one can arbitrarily increase the measurement accuracy by choosing an appropriate measurement device, specifically by reducing $ a $, simply increasing the reading accuracy of the pointer cannot improve above a certain value the precision of a simultaneous measurement of  the quantity  $ J $ and its speed-of-variation owing to the \so{Brown}ian motion of the measuring instrument. One can thus reduce $ a $ by reinforcing the magnetic field in a moving coil galvanometer with given mechanical properties and as a consequence  enhance the accuracy of current measurement at least in principle arbitrarily;     one cannot, however, achieve any reduction of the product $\Delta J\,\Delta\dot{J}  $ by a simple increase of the reading accuracy for example by magnifying the deflection using a microscopic reading pointer~\footnote{G. Ising, \textit{Ann. d. Phys.} \textbf{14} 755, 1932.}) 
 or using a thermal relay~\footnote{N. Moll u. N. Burger, \textit{Phil. Mag.} \textbf{1,} 624, 1925} 
  or a light electric relay~\footnote{L. Bergmann, \textit{Phys,ZS.} \textbf{32,} 688, 1931.})  
 
 The problem of the limits of measurement accuracy due to \so{Brown}ian motion of instruments, in particular galvanometers, has been recently repeatedly discussed by several authors~\footnote{G. Ising, \textit{Phil. Mag.} \textbf{1,} 827, 1926;  \textit{Ann.d.Phys.} \textbf{8,} 911, 1931;  \textbf{14,} 755,	1932;   F. Zernike, \textit{ZS. f. Phys.} \textbf{40,} 628, 1926;  \textbf{79,} 516, 1932;    R. Gans,   \textit{Schriften d. K{\"o}nigsberger Gel. Ges.} \textbf{7,} 177, 1930;   M. Czerny, \textit{Ann. d. Phys.} 12, 993, 1932}
 and it has been thoroughly discussed with which procedures one can perform the most accurate possible measurement of a quantity of interest with an instrument of a given type. In my opinion, these discussion have always overlooked an important point. The task of the experimentalist is certainly that of recording the quantity $ J $ of interest as a function of time, i.e. the function $ J(t) $ with the highest accuracy  possible.
 If one restricts [the attention] to a short interval of time, this requirement is equivalent to the task of \emph{determining a quantity $ J $ and its variation speed $ \dot{J} $ at a given instant of time with the highest possible accuracy}. The relation (\ref{5:general}) teaches that with a given instrument  this is possible only with a certain uncertainty completely independent from any procedure to increase the reading accuracy of the pointer.
 
 The procedure suggested by many authors to increase the maesurement accuracy of $ J $ despite the \so{Brown}ian motion by taking many readings and taking their average which should be then more precise than an individual measurement or by using an integrating measuring instrument makes sense only when one knows in advance that the quantity of interest is exactly constant. But how can one know this without having performed first a corresponding measurement to ascertain such stipulation? If one really tries this, then one would obtain by repeated observation or by continuous recording a time dependence of the [pointer] deflection (because of the \so{Brown}ian motion) whence it is certainly not possible to determine whether the observed quantity remains constant or whether it varies in time within the limit of accuracy of the recorded oscillations.
 This circulus vitiosus is the reason why the method proposed to increase measurement accuracy is not really feasible.
 
Actually we can even say with certainty that the requirement of constancy of $ J $ implied by the mentioned procedure is certainly not satisfied because any macroscopically defined quantity, which can be measured by a macroscopic measurement instrument, undergoes oscillations. For instance, in reality there is certainly no constant electromotive force even if the power source is protected from external interference with all possible refinement  because of  the occurrence of spontaneous potential oscillations induced by the thermal motion of electrons how it has been experimentally shown by several researchers over the last years~\footnote{J. B. Johnson, \textit{Phys. Rev.} \textbf{29,} 367, 1927;  \textbf{32,} 97, 1928;   N. H. Williams, \textit{ibidem}   \textbf{40,} 121, 1932;   L. S. Ornstein,  H. C. Burger, J. Taylor and W. Clarkson, \textit{Proc. Roy. Soc. London (A)} \textbf{115,} 391, 1927.}.
Thus to measure an electromotive force with the highest possible accuracy obviously means  to record as precisely as possible its time dependence or in a short time interval to simultaneously measure as precisely as possible the electromotive force and its variation velocity. But, as we have shown above, this accuracy has because of the \so{Brown}ian motion an upper limit which is independent of the way the measurement is performed.
 
 \section{}\label{sec:6}
 The results reported in the previous paragraphs are, as it has been repeatedly mentioned, due to the formal analogy between the fundamental differential equations of classical diffusion theory and quantum mechanics, a fact which becomes particularly evident when contrasting the equations (\ref{freediff}) and (\ref{freeS}) of \S~1. Already there we have however pointed out essential formal differences between the two equations. We now want to try to understand the physical origins of these differences. The following considerations should at the same time contribute to clarify certain ambiguities, which have recently been highlighted by \so{Ehrenfest} with the invitation to the physicists to tackle these problems.
 
 Classical diffusion can be regarded as a current which, as we saw in \S~1, is governed by a differential equation of the form (\ref{freediff}), where $ F $ is a real differential operator and $ u $ is a real function of position and time, representing the density of the diffusing element. It follows that it must be possible from the assignement of $ u $ at any instant of time to compute the density distribution 
 at any later (and of course also earlier) instant of time. In contrast to a problem of ordinary hydrodynamics, the diffusion current in the system under consideration is thus completely determined by the assignement at an arbitrary instant of time of the density as a function of the coordinates, without simultaneously requiring the knowledge of current velocity as a function of the coordinates. 
 This is due to the fact that the current velocity defined by equation (\ref{3:cv1d}) is a function of $ u $ and the coordinates alone and does not depend on the history of the system. Thus if $u(x, y, z)$ is known, then it also specifies $ v(x, y, z) $ and therefore the evolution of the system in the following time step is completely determined in the sense of classic hydrodynamics.
 
 We also note that a time reversal operation, an exchange of $ t $ with $ -t $ in equation (\ref{freediff}) is not possible because $ D $,
 the diffusion coefficient, owing to its molecular theoretical meaning, is positive-definite.
 The diffusion process is therefore ``irreversible''. This is also evident from the fact that the velocity current is for given $ u $ a pure function of the position, so the initial velocities are not reversible and are determined solely by the collisions with the surrounding molecules.
 
 The situation is quite different in the quantum mechanical case.
 Since the particle motion is not disturbed here by collisions with the molecules of the surrounding element, the motion of the particle cluster is essentially determined by the initial positions and the initial velocities of the particles. It is therefore clear that there cannot be a differential equation for the density function $ w $ in the same way as it occurs for classic diffusion. That on the contrary an equation of the form (\ref{freeS}) holds can be most easily seen from the point of view of wave mechanics. From this point of view, the particle cluster forms a ``wave packet'', i.e., a superposition of harmonic partial waves of the form
 \begin{align}
 \psi_{k}=\varphi_{k}e^{2\,\pi\, i \,E_{k}\,t/h}
 \nonumber
 \end{align}
 the number whereof has the cardinality of the continuum for the boundary conditions considered here. Here $ \varphi_{k} $ stands for the ``\emph{amplitude function}'' a complex function of the position of the form
 \begin{align}
 \varphi_{k}=a_{k}e^{ i \,S_{k}}
 \nonumber
 \end{align}
 containing two real functions of the position, the amplitude $ A_{k} $ and the phase $ S_{k} $. The assignment of all the $ A_{k} $'s and $ S_{k} $'s as functions of the position fully specifies the $ \varphi $ in the wave packet under consideration at a given instant time as well as for every later (or earlier) instants of time in consequence of the differential equation (\ref{freeS}), which is physically obvious, since the fate of each partial wave is determined by the specification of amplitude and phase at time zero and thus also the fate of the wave packet created by interference from the partial waves. So it is immediately comprehensible that for description of the state of the wave field two scalars or one complex function, the \so{Schr\"odinger} function, are necessary.
 
 Since the density of the cluster under consideration (now considered from the corpuscular point of view) is specified solely by $ |\psi| $ according to equation (\ref{Born}), the assignement of $ \psi $ as a function of the position entails more detailed information than the distribution of the particles' \emph{positions} at a certain instant of time.  According to what said above, as the fate of the cluster is determined by $ \psi $, it is evident that the assignement of $ \psi $ contains information also about the distribution of the \emph{velocities} at a certain instant of time. 
 If, conversely, the initial velocities are not known, then it is not possible from the initial distribution alone to
 make predictions about the motion of the particles' cluster. In fact there cannot be a differential equation for $ |\psi| $. Nevertheless 
 only the density $ w=\psi\psi^{*} $ or, interpreted as a virtual entity, the probability density of the position, is observable and not $ \psi $
 itself. This paradoxical state of the matter can be immediately explained as a consequence of the uncertainty relations. Were $ \psi $ indeed observable then according to our discussion the position and velocity distribution would be simultaneously assigned for our particle cluster which is not possible!
 
The fact that the coefficient on the left side of equation (\ref{freeS}) must be purely imaginary or the diffusion coefficient $ \epsilon $ in (\ref{sc}) must be purely imaginary can be seen as follows: if at an arbitrary instant of time the phases $ S_{k} $ of all the partial waves are reversed by $ 180^{\circ} $, then every $ \phi_{k} $ turns into $ \phi_{k}^{*} $ and therefore $ \psi $ into $ \psi^{*} $. At the same time, however, the reversal of all phases means turning all wave processes in the opposite direction or the complete reversal of the motion of the wave packet.
The exchange of $ \psi $ with its conjugate complex value $ \psi^{*} $ means nothing else than a time reversal, and the differential equation (\ref{freeS}), which $ \psi $ satisfies , must therefore remain unchanged under the simultaneous replacement of $ \psi $ with $ \psi^{*} $ and of $ t $ with $ -t $. This is actually only possible, provided that the \so{Hamilton} operator $ H $ is time independent, if the coefficient of $ \frac{\partial \psi}{\partial t} $ is purely imaginary. The occurrence of the imaginary diffusion coefficient means, as \so{Schr\"odinger} has already pointed out~\footnote{Erwin \so{Schr{\"o}dinger},~\textit{loc. cit.}, ref~5.}, simply the reversibility of the quantum mechanical ``diffusion'' in contrast to the classical one, a discrepancy that was already emphasized in \S~2 and 3 in the [discussion of the] differences between equations (\ref{quadratic}) and (\ref{3:quadratic}).

\vspace{0.5cm}

Prague, January 1933.

\appendix

\section{Translation notes}

\subsection{Translation style} We tried to reproduce the style of the original prose by not splitting the long(!) sentences. We used square brackets $ [\dots] $ for text added either to maintain syntactic congruence or to emphasize the meaning implicit in the construction of the original sentence.

\subsection{Sources} Some references are not exact. We could not retrieve in particular N. L. Williams's paper. Furthermore we could not find any paper on Physical Review by J. B. Johnson in 1927. Notice that the references that Fürth displays in footnotes, as costumary in the original journal, have been repeated at the end of the present work.

\subsection{Influence on stochastic mechanics:} F\"urth's paper is discussed in F\'enyes \cite{Fen52}. This paper lays down the foundation of what will be Nelson's ``stochastic mechanics'' program~\cite{Nelson85}.
In the introduction of~\cite{Fen52} Fe\'nyes states:
\begin{quote}
	\emph{Although \so{F\"urth} has demonstrated  the existence in diffusion theory of a relation that is formally analogous to \so{Heisenberg}'s, in his opinion the two relations cannot have the same meaning, because the \so{Fokker} equation cannot be valid in quantum mechanics.}
\end{quote}
 Fe\'nyes's findings  in~\cite{Fen52} as summarized in the paper's abstract are:
 \begin{quote}
 	\emph{There are also certain uncertainty relations for \so{Markov} processes. A certain probability-amplitude function can also be assigned to a \so{Markov} processes. The \so{Fokker} equation is also valid in quantum mechanics. The \so{Heisenberg} relation is a special case of the uncertainty relation of the \so{Markov} processes. The wave-mechanical wave function is a special case of probability-amplitude functions governed by \so{Markov} processes.  The wave-mechanical processes are special \so{Markov} processes. The \so{Heisenberg} relation is (in contrast to the previous interpretation) exclusively a consequence of the statistical approach, and is independent of the disturbances occurring in the two measurements.}
 \end{quote}
Finally F\"urth and F\'enyes are known in the stochastic quantization community where are somewhat considered as precursors of the Parisi-Wu method (see e.g, the discussion in the introduction of ~\cite{Nam92}).

\subsection{General derivation of the uncertainty relation} Here we reproduce for convenience Fe\'nyes' argument. To start with, we recall the diffusion pathwise probabilistic definition of current velocity and its relation with the ``coefficients'' of the Fokker-Planck equation.

Let us consider a stochastic process $ \left\{ \bm{\xi}_{t} \right\}_{t\,\geq\,0} $ with drift
\begin{align}
\bm{b}(\bm{x},t)=\lim_{s\,\searrow\,0}\operatorname{E}\left(\frac{\bm{\xi}_{t+s}-\bm{\xi}_{t}}{s}\bigg{|}\bm{\xi}_{t}=\bm{x}\right)
\nonumber
\end{align}
and diffusion
\begin{align}
\mathds{D}(\bm{x},t)=\lim_{s\,\searrow\,0}\operatorname{E}\left(\frac{(\bm{\xi}_{t+s}-\bm{\xi}_{t})\,\otimes\,(\bm{\xi}_{t+s}-\bm{\xi}_{t})}{s}\bigg{|}\bm{\xi}_{t}=\bm{x}\right)
\nonumber
\end{align}
We assume that drift and diffusion enjoy regularity properties in $ \mathbb{R}^{d} $ such that the transition probability density $ \mathcal{T} $ satisfies Kolmogorov's forward (Fokker-Planck)
\begin{equation}
\begin{split}
&\partial_{t}\mathcal{T}(\bm{x},t | \bm{y},s) +\partial_{\bm{x}}\cdot\bm{b}(\bm{x},t)\mathcal{T}(\bm{x},t| \bm{y},s)\\
&\qquad = \dfrac{1}{2}\operatorname{Tr} \partial_{\bm{x}}\,\otimes\,\partial_{\bm{x}}\mathds{D}(\bm{x},t)\mathcal{T}(\bm{x},t| \bm{y},s)
\label{ap:Kf}
\end{split}
\end{equation}
and backward equations
\begin{equation}
\begin{split}
&\partial_{s}\mathcal{T}(\bm{x},t\big{|}\bm{y},s)+\bm{b}(\bm{y},s)\cdot\partial_{\bm{y}}\mathcal{T}(\bm{x},t\big{|}\bm{y},s)\\
&\qquad{}+ \frac{1}{2}\operatorname{Tr}\mathds{D}(\bm{y},s)\partial_{\bm{y}}\,\otimes\,\partial_{\bm{y}}\mathcal{T}(\bm{x},t\big{|}\bm{y},s)=0
\label{ap:Kb}
\end{split}
\end{equation}
both subject to the time boundary condition 
\begin{align}
\lim_{t-s\,\searrow\,0}\mathcal{T}(\bm{x},t\big{|}\bm{y},s)=\delta^{d}(\bm{x}-\bm{y})
\nonumber
\end{align}
for any $ \bm{x},\bm{y} \,\in\,\mathbb{R}^{d}$, and $ t\,\geq\,s\,\geq\,0 $. In such a case, the probability density of the process $ \left\{ \bm{\xi}_{t} \right\}_{t\,\geq\,0} $ 
evolving from any reasonable initial data $  \mathcal{p}_{\iota}(\bm{x})$ also satisfies the Fokker-Planck equation by means of the Markov property
\begin{align}
\mathcal{p}(\bm{x},t)=\int_{\mathbb{R}^{d}}\mathrm{d}^{d}\bm{y}\,\mathcal{T}(\bm{x},t\big{|}\bm{y},s)\mathcal{p}_{\iota}(\bm{y},s)
\qquad \forall\,t\,\geq\,s\,\geq\,0 
\nonumber
\end{align}
Under these hypotheses, the current velocity is defined as the conditional expectation of the time symmetric increment (see e.g. \cite{Nelson85})
\begin{align}
\bm{v}(\bm{x},t)=\lim_{s\,\searrow\,0}\operatorname{E}\left(\frac{\bm{\xi}_{t+s}-\bm{\xi}_{t-s}}{2\,s}\bigg{|}\bm{\xi}_{t}=\bm{x}\right)
\nonumber
\end{align}
The evaluation the conditional expectation yields 
\begin{align}
\bm{v}(\bm{x},t)=
\bm{b}(\bm{x},t)-\frac{1}{2\,\mathcal{p}(\bm{x},t)}\partial_{\bm{x}}\mathds{D}(\bm{x},t)\mathcal{p}(\bm{x},t)
\label{ap:cv}
\end{align}
where $ \mathcal{p} $ is the probability density of $ \left\{ \bm{\xi}_{t} \right\}_{t\,\geq\,0} $.
The derivation of  (\ref{ap:cv}) may use Kolmogorov's \emph{time reversal relation} between the  probability and the forward $ \mathcal{T} $ and backward $ \mathcal{T}_{R} $ transition probability densities 
\begin{align}
\mathcal{T}_{R}(\bm{y},t\big{|}\bm{x},t+s)\mathcal{p}(\bm{x},t+s)=\mathcal{T}(\bm{x},t+s\big{|}\bm{y},t)\mathcal{p}(\bm{y},t)
\nonumber
\end{align}
eq.~(8) of Kolmogorov's 1937 paper \cite{Ko37}.  Kolmogorov's paper was known to F\'enyes but, obviously, could not be to F\"urth. Namely, under our working hypotheses the identity
\begin{align}
\operatorname{E}\left(\bm{\xi}_{t-s}\big{|}\bm{\xi}_{t}=\bm{x}\right)=\int_{\mathds{R}^{d}}\mathrm{d}^{d}y\,\bm{y}\,
\mathcal{T}_{R}(\bm{y},t-s\big{|}\bm{x},t)\nonumber\\
=\int_{\mathds{R}^{d}}\mathrm{d}^{d}y\,\bm{y}\frac{\mathcal{T}(\bm{x},t\big{|}\bm{y},t-s)\mathcal{p}(\bm{y},t-s)}{\mathcal{p}(\bm{x},t)}
\nonumber
\end{align}
holds true. We then obtain (\ref{ap:cv}) observing that as functions of $ \bm{y} $ the density obeys the Fokker-Planck 
equation whereas the transition probability satisfies Kolmogorov's backward equation (\ref{ap:Kb}).

An important consequence of (\ref{ap:cv}) for the line of reasoning of F\'enyes is that the Fokker-Planck equation satisfied by a \emph{given density} $ \mathcal{p}(\bm{x},t) $  once expressed in terms of the current velocity takes the form of  a mass continuity equation
\begin{align}
\partial_{t}\mathcal{p}(\bm{x},t)+\partial_{\bm{x}}\cdot\bm{v}(\bm{x},t)\mathcal{p}(\bm{x},t)=0
\nonumber
\end{align}
F\'enyes' uncertainty relation  follows by applying Cauchy-Schwarz inequality to the product of the variances of respectively the position and current velocity processes of  $ \left\{ \bm{\xi}_{t} \right\}_{t\,\geq\,0} $:
\begin{align}
\operatorname{V}\bm{\xi}_{t}=\int_{\mathbb{R}^{d}}\mathrm{d}^{d}\bm{x}\mathcal{p}(\bm{x},t) \left\|\bm{x}-\operatorname{E}\bm{\xi}_{t}\right\|^{2}
\nonumber
\end{align}
and
\begin{align}
\operatorname{V}\bm{v}(\bm{\xi}_{t},t)=\int_{\mathbb{R}^{d}}\mathrm{d}^{d}\bm{x}\mathcal{p}(\bm{x},t) \left\|\bm{v}(\bm{x},t)-\operatorname{E}\bm{v}(\bm{\xi}_{t},t)\right\|^{2}
\nonumber
\end{align}
Cauchy-Schwarz immediately yields
\begin{align}
&\left(\operatorname{V}\bm{\xi}_{t}\right)\,\operatorname{V}\bm{v}(\bm{\xi}_{t},t)\,\geq\,
\left |\int_{\mathbb{R}^{d}}\mathrm{d}^{d}\bm{x}\mathcal{p}(\bm{x},t) \left(\bm{x}-\operatorname{E}\bm{\xi}_{t}\right)\right.\nonumber\\
&\qquad \left.{}\cdot\left(\bm{v}(\bm{x},t)-\operatorname{E}\bm{v}(\bm{\xi}_{t},t)\right)\right |^{2}
\nonumber
\end{align}
We then use (supposing that the density vanishes at infinity sufficiently fast)
\begin{align}
\operatorname{E}\bm{v}(\bm{\xi}_{t},t)=\operatorname{E}\bm{b}(\bm{\xi}_{t},t)
\nonumber
\end{align}
to write
\begin{equation*}
\begin{split}
&\int_{\mathbb{R}^{d}}\mathrm{d}^{d}\bm{x}\mathcal{p}(\bm{x},t) \left(\bm{x}-\operatorname{E}\bm{\xi}_{t}\right)\cdot\left(\bm{v}(\bm{x},t)-\operatorname{E}\bm{v}(\bm{\xi}_{t},t)\right)=
\nonumber\\
&\hspace{0.5cm}
\int_{\mathbb{R}^{d}}\mathrm{d}^{d}\bm{x}\mathcal{p}(\bm{x},t) \left(\bm{x}-\operatorname{E}\bm{\xi}_{t}\right)\\&\qquad\qquad{}\cdot\left(\bm{b}(\bm{x},t)-\operatorname{E}\bm{b}(\bm{\xi}_{t},t)-\frac{1}{2\,\mathcal{p}(\bm{x},t)}\partial_{\bm{x}}\mathds{D}(\bm{x},t)\mathcal{p}(\bm{x},t)\right)
\end{split}
\end{equation*}
We thus obtain Fe\'nyes's inequality 
\begin{equation}
\begin{split}
&\left(\operatorname{V}\bm{\xi}_{t}\right)\,\operatorname{V}\bm{v}(\bm{\xi}_{t},t)
\,\geq\,
\left |
\operatorname{E}\big{(}\bm{\xi}_{t}\cdot\bm{b}(\bm{\xi}_{t},t)\big{)}\phantom{\frac12}\right.\\
&\qquad\left.-\left(\operatorname{E}\bm{\xi}_{t}\right)
\cdot\operatorname{E}\bm{b}(\bm{\xi}_{t},t)+\frac{1}{2}\operatorname{E}\operatorname{Tr}\mathds{D}(\bm{\xi}_{t},t)
\right |^{2}
\label{ap:Fenyes}
\end{split}
\end{equation}
We observe that
\begin{itemize}
	\item the bound reduces to F\"urth's whenever the drift is negligible (setting e.g. $ \bm{b}=0 $) or whenever the process is positively 
	correlated with the drift.
	\item At equilibrium $ \bm{v}=0 $ by definition. In general therefore we can only expect
	\begin{align}
	\left(\operatorname{V}\bm{\xi}_{t}\right)\,\operatorname{V}\bm{v}(\bm{\xi}_{t},t)
	\,\geq\,
	0
	\nonumber
	\end{align}
\end{itemize}
To substantiate the last observation we consider the case
\begin{align}
\bm{b}(\bm{x})=-\partial_{\bm{x}}U(\bm{x})
\hspace{0.5cm}\&\hspace{0.5cm}
\mathds{D}=D\,\mathds{1}
\nonumber
\end{align}
Then equilibrium means (assuming $ U $ positive definite and confining)
\begin{align}
\mathcal{p}(\bm{x})\propto e^{-\frac{2\,U(\bm{x})}{D}}
\nonumber
\end{align}
so that
\begin{align}
&\int_{\mathbb{R}^{d}}\mathrm{d}^{d}x\,\mathcal{p}(\bm{x})\,\bm{x}\cdot\bm{b}(\bm{x})
=
-\int_{\mathbb{R}^{d}}\mathrm{d}^{d}x\,\mathcal{p}(\bm{x})\,\bm{x}\cdot\partial_{\bm{x}}U(\bm{x})\nonumber\\
&\qquad{}=\frac{D}{2}\int_{\mathbb{R}^{d}}\mathrm{d}^{d}x\,\bm{x}\cdot\partial_{\bm{x}}\mathcal{p}(\bm{x})=-\frac{D}{2}d
\nonumber
\end{align}
whereas
\begin{align}
\int_{\mathbb{R}^{d}}\mathrm{d}^{d}x\,\mathcal{p}(\bm{x})\,\bm{b}(\bm{x})=\frac{D}{2}\int_{\mathbb{R}^{d}}\mathrm{d}^{d}x\,\partial_{\bm{x}}\mathcal{p}(\bm{x})=0
\nonumber
\end{align}
Thus the right hand side of (\ref{ap:Fenyes}) vanishes.

Finally we notice, see e.g. \cite{PePi20}, that when $ \mathds{D}=D\,\mathds{1} $
\begin{align}
\mathcal{S}_{\mathcal{t}_{\mathcal{f}},\mathcal{t}_{\mathcal{i}}}=\operatorname{E}\int_{\mathcal{t}_{\mathcal{i}}}^{\mathcal{t}_{\mathcal{f}}}
\mathrm{d}t \,\|\bm{v}(\bm{\xi}_{t},t)\|^{2}
\nonumber
\end{align}
is proportional to the average entropy production of by the process $ \left\{ \bm{\xi}_{t} \right\}_{t\in [\mathcal{t}_{\mathcal{f}},\mathcal{t}_{\mathcal{i}}]} $ a fact which exhibits the interest of F\"urth-F\'eynes uncertainty relations for contemporary developments in stochastic thermodynamics. 

\subsection{Continuity of Brownian motion} A qualitative understanding of the result based on the present-day theory of Brownian motion is as follows. Paths of a Brownian motion are with probability one  
H\"older continuous with exponent $ 1/2 $. This means that they are nowhere differentiable. As a consequence if one observes a Brownian particle with increasing resolution will magnify the resolution of the
particle position but at the same time register an increase without bound of derivatives of the trajectory.

\subsection{Acknowledgement to K.~L\"owner}

It is intriguing to read that F\"urth acknowledges K.~L\"owner for hints in the derivation of the quantum uncertainty relation. Karel L\"owner, also known as Charles Loewner after emigration to the U.S., was a mathematician whose work on conformal mappings led to the discovery of what is now commonly known as the ``Loewner differential equation''. The stochastic extension of his work is the stochastic Loewner equation or Schramm-Loewner evolution (SLE) a family Markov processes describing interfaces in two dimensional critical systems. The study of SLE has attracted a lot of attention both in the physics and mathematics community over the last two decades see e.g. \cite{BaBe06,Kem17}.

\end{document}